\documentclass[twocolumn,11pt]{aastex631}
\usepackage{graphicx,amsmath}
\usepackage{color}
\usepackage{booktabs}
\usepackage{bm}

\usepackage{times}
\usepackage{float}

\usepackage{ulem}
\usepackage{multirow}
\usepackage{amsmath}

\newcommand{\adsurl}[1]{\href{#1}{ADS}}
\providecommand{\url}[1]{\href{#1}{#1}}

\newcommand{\be}{\begin{equation}}
\newcommand{\ee}{\end{equation}}
\newcommand{\bea}{\begin{eqnarray}}
\newcommand{\eea}{\end{eqnarray}}

\usepackage{color}
\usepackage{ifthen}
\newboolean{editorial}
\setboolean{editorial}{true}
\newcommand{\editorial}[2]{\ifthenelse{\boolean{editorial}}{\textcolor{red}{[\textsf{{{#1}}}: }\textcolor{blue}{\textsf{{#2}}}\textcolor{red}{]}}{}}

\shorttitle{Lensing rates of sBBHs produced by dynamical channel}
\shortauthors{Chen}

\begin{document}

\title{
Event rate of strongly lensed gravitational waves of stellar binary black hole mergers produced by dynamical interactions
}

\correspondingauthor{Zhiwei Chen}
\email{chenzhiwei171@mails.ucas.ac.cn}

\author[0000-0001-7952-7945]{Zhiwei Chen}
\affiliation{National Astronomical Observatories, Chinese Academy of Sciences, 20A Datun Road, Beijing 100101, China}
\affiliation{School of Astronomy and Space Sciences, University of Chinese Academy of Sciences, 19A Yuquan Road, Beijing 100049, China}

\begin{abstract}
Gravitational waves emitted from stellar binary black hole (sBBH) mergers can be {gravitationally lensed} by intervening galaxies and detected by future ground-based detectors. A large amount of effort has been put into the estimation of the detection rate of lensed sBBH originating from the {evolution of massive binary stars} (EMBS channel).  However, sBBHs produced by the dynamical interaction in dense clusters (dynamical channel) may also be dominant in our universe and their intrinsic distribution of physical properties {can  be significantly different} from those produced by massive stars, especially mass and redshift distribution.  {In this paper, we investigate the event rate of lensed sBBHs produced via dynamical channel by Monte Carlo simulations and the number is {$16_{-12}^{+4.7} $ $\rm yr^{-1}$ for the Einstein telescope and $24_{-17}^{+6.8}$ $ \rm yr^{-1}$ for Cosmic Explorer}, of which the median is about $\sim 2$ times the rate of sBBHs originated from EMBS channel (calibrated by the local merger rate density estimated for the dynamical and the EMBS channel, i.e., $\sim 14_{-10}^{+4.0}$ and $19_{-3.0}^{+42} \rm Gpc^{-3}yr^{-1}$ respectively)}. Therefore, one may  {constrain} the fraction of both EMBS and dynamical channels through the comparison of the predicted and observed number of lensed sBBH events  {statistically}.  

\end{abstract}



\section{Introduction}
\label{sec:intro}

Stellar binary black hole (sBBH) merger is one of the most important gravitational wave (GW) sources in the universe. It has been studied intensively in the literature, especially since the first detection of GW150914 and other subsequent observations by Laser Interferometer GW Observatories (LIGO) and VIRGO \citep{2016PhRvL.116m1102A, 2019PhRvX...9c1040A, 2020arXiv201014527A, 2021arXiv211103606T, 2021arXiv211103634T}. In principle, unlike binary neutron stars, sBBHs are not likely to have detectable electromagnetic (EM) counterparts at the same time. Thus, GW detection 
maybe the only way to study their origin and physical properties. By the following third-generation detectors with increased sensitivities, such as Einstein Telescope (ET) and Cosmic Explorer (CE), one may expect to detect $10^4-10^5$ sBBHs events per year to a very deep redshift. 
A small fraction ($10^{-3}-10^{-4}$) of these events could be lensed by intervening galaxies {\citep[e.g.,][]{2010MNRAS.405.2579O,2018MNRAS.476.2220L,2022ApJ...929....9X}}. 

Gravitational lensing of GWs has been studied in great detail {\citep[e.g.,][]{1996PhRvL..77.2875W,1998PhRvL..80.1138N,2003ApJ...595.1039T,2018PhRvD..98j4029D,2019A&A...627A.130D,2022PhRvD.106b3018G, 2023PhRvD.107d3029C,2023arXiv230404800L}. }
 {In the geometrical optics range}, GWs emitted by sBBH mergers may be deflected by intervening galaxies and produce  multiple images with different magnification factors. Among these images, there exist time-delays due to their different optical path and gravitational potential.  These special events are unique and powerful probes to constrain cosmological parameters for their precision on the measurement of time-delays between different images, provided their positions in (or associated with) the host galaxies are known \citep[e.g.,][]{2017NatCo...8.1148L, 2019ApJ...873...37L,2020MNRAS.498.3395H}. {Moreover, if the associated lensed host galaxies can be identified with future sky-surveys, it is also possible to constrain the relative position of these unique lensed sBBH mergers in their host galaxies \citep[e.g.,][]{2020MNRAS.497..204Y,2020MNRAS.498.3395H,2022arXiv220408732W}. Thus, the detection rate of lensed sBBH having identifiable lensed host galaxies are of great significance.}

As shown in \citet{2022ApJ...940...17C}, the detection fraction of the host galaxies to the lensed GW events can be {different} if sBBHs originated from different formation mechanisms. The fraction is $\sim 20\%$ assuming optimistic limiting magnitude for the evolution of massive binary stars \citep[hereafter denoted as the EMBS channel, e.g.,][]{2016Natur.534..512B, 2018MNRAS.480.2011G,2019MNRAS.486.2494G}. As for sBBHs produced by the dynamical interactions in dense stellar systems \citep[hereafter the dynamical channel, e.g.,][]{1993Natur.364..423S, 2000ApJ...528L..17P, 2016ApJ...824L...8R,2022MNRAS.511.5797M}, this fraction is slightly lower, i.e., $\sim 15\%$. By multiplying this fraction with the predicted detection rate of lensed sBBHs originating from different formation channels, one may estimate the detection rate of both lensed events with future sky-surveys and therefore constrain the origin of sBBHs.  

Most of the works done so far focus on predicting the detection rate of lensed sBBHs produced by EMBS channel (or simply follow the star formation rate, SFR) \citep[e.g.,][]{ 2014JCAP...10..080B, Piorkowska:2013eww, 2015JCAP...12..006D, 2018MNRAS.476.2220L, 2019ApJ...874..139Y, PhysRevD.103.104055, 2021MNRAS.501.2451M, 2021ApJ...921..154W,2022MNRAS.509.3772Y}. However, as proposed in \citet{2021RNAAS...5...19R}, dynamical interactions in globular clusters can also dominate the formation of sBBHs. We note that sBBHs produced by the dynamical channel are likely to have larger masses and luminosity distances, which may result in a more considerable lensed rate rather than sBBHs produced by the EMBS channel. This may enhance the identification between the EMBS channel and dynamical channel directly by their different detection rate of lensed sBBH mergers.

In this paper, we estimate the detection number per year of lensed sBBHs originated from dynamical interactions in dense clusters with third-generation detectors and their networks, i.e., {ET} \footnote{ET-D design \citep[][]{Hild_2011} \href{http://www.et-gw.eu/}{http://www.et-gw.eu/} } and {CE} \footnote{Stage-2 phase \citep[][]{2019BAAS...51g..35R} \href{https://cosmicexplorer.org/}{https://cosmicexplorer.org/}}, by adopting SIE (singular isothermal ellipsoid)+ external shear model \citep[e.g.,][]{1991ApJ...373..354K,1997MNRAS.291..211W, 1998ApJ...495..157K} as the lens model and applying more realistic templates \citep[e.g.,][]{2019PASP..131b4503B,2019PhRvD.100b4059K} to generate GW waveforms rather than the approach developed by \citet{1996PhRvD..53.2878F}. {We then show that by multiplying the detection fraction proposed in \citet{2022ApJ...940...17C} , the detection rate of lensed sBBH having identifiable lensed host galaxies can be different for those produced by the dynamical channel and the EMBS channel, which may offer an opportunity to constrain the origin of sBBH mergers. } 

This paper is organized as follows. In Section~\ref{sec:method}, we briefly introduce the method to estimate the detection rate of lensed sBBH produced by the dynamical channel.  In Section~\ref{sec:results}, we present our main results. Discussions and conclusions are given in Section~ \ref{sec:con}. Throughout the paper, we adopt the cosmological parameters as $(h_0,\Omega_{\rm m},\Omega_\Lambda)=(0.68,0.31,0.69)$ \citep{Aghanim2020}.

\section{methodology}
\label{sec:method}

In this section, we first introduce the intrinsic distribution of sBBH formed by dynamical interactions in dense clusters (Sec \ref{subsec:rate}). Then we show the signal detectability of third-generation gravitational wave detectors (Sec \ref{subsec:detect}) and lensing statistics adopting 
SIE+ external shear model (Sec \ref{subsec:statistics}) . A detailed description of the Monte Carlo method is illustrated in Sec \ref{sub:MC}. 

\subsection{Intrinsic sBBH rates}
\label{subsec:rate}
The number density distribution of the sBBH merger GW events produced by the dynamical channel can be described as
\begin{equation}
\frac{d\dot{N}}{d\boldsymbol{m_1}dq dz}=\frac{\boldsymbol{R_D}(z,\boldsymbol{m_1},q)}{1+z} \frac{dV(z)}{dz}  
\label{source}
\end{equation}

where $m_1$ is the primary mass, $q$ is the mass ratio, and $\boldsymbol{R_D}(z,\boldsymbol{m_1},q)$ is the merger rate density of the dynamical channel with the primary mass in the range from \boldsymbol{$m_1$} to at \boldsymbol{$m_1+ dm_1$}, the mass ratio in the range from $q$ to at $q+dq$ at redshift $z$. The factor $1/(1+z)$ accounts for the time dilation. 
%

The merger rate density $\boldsymbol{R_D}(z,\boldsymbol{m_1},q)$ can be estimated by using both dynamical simulations on the formation of sBBHs and simple descriptions on the formation and evolution of globular clusters \citep[][]{2021MNRAS.500.1421Z}, 
\begin{equation}
\begin{aligned}
\boldsymbol{R_{\rm D}}(z,m_1,q)= &\left.\iiint \frac{\dot{M}_{\mathrm{GC}}}{d \log _{10} M_{\mathrm{Halo}}}\right|_{z(\tau)} \frac{1}{\left\langle M_{\mathrm{GC}}\right\rangle} P\left(M_{\mathrm{GC}}\right) \\
& \times R\left(r_{\mathrm{v}}, M_{\mathrm{GC}}, \tau-t(z)\right) d M_{\mathrm{Halo}} d M_{\mathrm{GC}} d \tau,
\end{aligned}
\end{equation}
where {the distribution of} $q$ is assumed to be proportional to $q$ in the
range from $0.5$ to $1$, $\frac{\dot{M}_{\mathrm{GC}}}{d \log _{10} M_{\mathrm{Halo}}}$ is the comoving SFR in globular clusters per galaxies of a given halo mass $M_{\rm Halo}$ at given redshift $z(\tau)$ (or a given formation time $\tau$ ), $P(M_{\rm GC})$ is the cluster initial mass function, ${\left\langle M_{\mathrm{GC}}\right\rangle}$ is the mean initial mass of a globular cluster and $R(r_{\rm v}, M_{\rm GC}, t)$ is the merger rate density of sBBHs in a globular cluster with initial virial radius $r_{\rm v}$ and mass $M_{\rm GC}$ at time $t(z_{\rm s})$. Here we adopt a log-normal distribution form for $\frac{\dot{M}_{\mathrm{GC}}}{d \log _{10} M_{\mathrm{Halo}}}$, which assumes $50\%$ of clusters form with $r_{\mathrm{v}} = 1$\,pc and $50\%$ form with $r_{\mathrm{v}} = 2$\,pc \citep{2018ApJ...866L...5R},
\begin{equation}
\left.\frac{\dot{M}_{\mathrm{GC}}}{d \log _{10} M_{\text {Halo}}}\right|_{z} \approx \frac{A(z)}{\sqrt{2 \pi} \sigma(z)}\exp \left(-\frac{\log _{10} M_{\text {Halo}}-\mu(z)}{2 \sigma(z)^{2}}\right), 
\end{equation}
where $A(z)$, $\mu(z)$, and $\sigma(z)$ are fitted polynomials in the redshift
$z$. More detailed descriptions of the estimates of  sBBH merger rate density via the dynamical channel can be found in \citet{2021MNRAS.500.1421Z}.

{In figure \ref{fig:tau}, the orange solid line shows the merger rate density evolution for dynamical channels, which is scaled by the median local merger rate density  $\sim 14\rm Gpc^{-3}yr^{-1}$ estimated by the \citet{2018ApJ...866L...5R}, while the orange shadow shows the evolution scaled by the corresponding lower and upper bound $\sim 4-18\rm Gpc^{-3}yr^{-1}$  {($90\%$ confidence interval)} \citep[e.g., ][]{2018ApJ...866L...5R}.} {For comparison, we also plot the results for the EMBS channel in a blue solid line, which is scaled by the median local merger rate density  $\sim 19\rm Gpc^{-3}yr^{-1}$ constrained by the first three observation runs of LIGO-Virgo-KAGRA \citep{2021arXiv211103606T, 2021arXiv211103634T}. The blue shadow represent the error range induced by the uncertainties of local merger rate density, $\sim 16-61\rm Gpc^{-3}yr^{-1}$  {($90\%$ confidence interval)} \citep[e.g., ][]{2021arXiv211103606T, 2021arXiv211103634T}.  The detailed calculation for this channel could be seen in \citet{2017cao}.  The median merger rate density from the EMBS channel peaks at $z_{\rm s}\sim 1.5$ and has a value of $\sim 80 \rm Gpc^{-3}yr^{-1}$, while those from the dynamical and channel peak at higher redshifts, i.e.,  $z_{\rm s}\sim 2.5$ and has a higher value of $\sim 106 \rm Gpc^{-3}yr^{-1}$. }

\subsection{GW detectability}
\label{subsec:detect}
 
The GW signal produced by sBBH mergers can be detected with the matched filtering method.  The signal-to-noise ratio ($\rm SNR$) for a single detector can be calculated by the self-inner product of the GW strain $h(f)$ with respect to the one-sided power spectrum of the GW detector $S_n(f)$ in the frequency domain,
\begin{equation}
     \rho^2=4\int_{f_{\rm min}}^{f_{\rm max}} \frac{|h(f)|^2}{S_n(f)}df
\label{eq:snr}
 \end{equation}
where ${f_{\rm min}}$ and ${f_{\rm max}}$ are the lower and upper limits of frequency of GW waveforms. As for the detection network, the total optimal SNR  can be simply calculated by the geometric mean of that of single detectors,
\begin{equation}
    \rho=\sqrt{\rho_{1}^2+\rho_{2}^2}
\end{equation}
where 1 and 2 denote for the first and second detectors (ET and CE in this paper) respectively.

{Instead of approximating the GW waveforms by the inspiral equations proposed by \citet{1996PhRvD..53.2878F}}, we employ the standard package {PyCBC} \citep[][]{2019PASP..131b4503B} to produce GW waveforms $h_{+}(f)$ and $h_{\times}(f)$ in the frequency domain for each sBBH merger originated from the dynamical channel.
We adopt the phenomenological model {IMRPhenomPv3} proposed by \citet{2019PhRvD.100b4059K}, which considers the dynamics of precessing binary black holes with two-spin effects. 
%
Then the total strain $h(f)$ received by GW detectors can be represented by:
\begin{equation}
    h(f)=F_{+}h_{+}(f)+F_{\times}h_{\times}(f),
\end{equation}
with the detector's antenna pattern function $F_{+}$ and $F_{\times}$:
\begin{equation}
\begin{aligned}
&F_{+} \equiv \frac{1}{2}\left(1+\cos ^{2} \theta\right) \cos 2 \phi \cos 2 \psi-\cos \theta \sin 2 \phi \sin 2 \psi \\
&F_{\times} \equiv \frac{1}{2}\left(1+\cos ^{2} \theta\right) \cos 2 \phi \sin 2 \psi+\cos \theta \sin 2 \phi \cos 2 \psi
\end{aligned}
\end{equation}
where $(\theta, \phi)$ is the spherical coordinates in the detector's frame, and $\psi$ is the orientation of sBBH towards the GW detector. 

The  GW signal is detectable once its $\rm SNR$ exceeds the threshold, i.e., $\rho>\rho_{0}$. As for the multiple images of lensed GW signals, the $\rm SNR$ could be magnified by gravitational lensing,
\begin{equation}
   \rho^{\rm len}=\sqrt{\mu}\rho
   \label{eq:mag}
   \end{equation}
where $\mu$ is the magnification factor of the lensed image. 

{In this paper, for the SIE+ external shear model \citep[e.g.,][]{1991ApJ...373..354K,1997MNRAS.291..211W, 1998ApJ...495..157K}, there are two main types of lensed images, i.e.,  double images case (the source locates within the outer critical line but outside the inner critical line) and quadruple images case (the source locates within both the outer and inner critical lines). } To ensure at least two lensed GW signals are detectable, we calculate the $\rm SNR$ for the fainter images of double images case and the {second} brightest images of quadruple images and check whether it is larger than the threshold adopted, i.e.,  $\rho_{0}=8$. {When one wish to detect four images, the $\rm SNR$ of the least brightest images in quadruple images case should excess the threshold $\rho_0$.}

\subsection{Lensing statistics}
\label{subsec:statistics}
Numerous observations have shown that the galaxy-galaxy strong lensing is dominated by elliptical galaxies  \citep[e.g.][]{1984ApJ...284....1T,2007MNRAS.379.1195M}. 
Therefore, the singular isothermal ellipsoid profile (SIE) with external shear is normally adopted as the lens model, by which most of the lensed events may produce either double or quadruple-lensed images \citep[e.g.][]{2010MNRAS.405.2579O,2018MNRAS.480.3842O,2018MNRAS.476.2220L}. {The optical depth $\tau(z_{\rm s})$, or the probability that a GW event can be lensed by intervening galaxies, is described as follows under the geometrical optics approximation} \citep[e.g.][]{2018MNRAS.476.2220L,2018MNRAS.480.3842O,2023MNRAS.518.6183M}.

\begin{equation}
\begin{aligned}
\tau(z_s)=&\frac{1}{4 \pi} \int_{0}^{z_s} \frac{dV({z_{\rm l}})}{dz_{\rm l}} dz_l \iint d\gamma_{1}d\gamma_{2}P_{\gamma}(\gamma_{1},\gamma_{2}) \int de P_{\rm e}(e) \\
& \times  \int \frac{dn(\sigma_{\rm v},z_{\rm l})}{d\sigma_{\rm v}} d\sigma_{\rm v} S_{\rm cr}(\sigma_{v},z_l,z_s,\gamma_{1},\gamma_{2},e) 
\end{aligned}
\label{eq:tau}
\end{equation}
where $z_{\rm l}$ is the redshift of  the lens, and $P_{\rm e}$ and $P_{\gamma}$ represent the probability distributions of the {axis-ratio} and the two-dimensional external shear in Cartesian coordinates, which describe the lens morphology and external environment near the line of sight. {Note here that in this paper, we assume the amplitude ${\gamma}$ follows a log-normal distribution with mean $\ln{ 0.05}$ and standard deviation $0.2$ and the direction of $\gamma$ is randomly distributed following \citet{2005ApJ...624...34H}. The axis-ratio distribution $P_{\rm e}$ is a truncated Gaussian distribution between $[0.2,1]$, with mean value $0.7$ and standard deviation $0.16$ consistent with the observations on early-type galaxies \citep{2003ApJ...594..225S}.}

The cross-section $S_{\rm cr}$ is dependent on the lens galaxy velocity dispersion, the redshifts of lens and source, eccentricity and external shear. We model the velocity distribution function (VDF) $dn(\sigma_{\rm v},z_{\rm l})/d\sigma_{\rm v}$ as a simple Schechter function given by \citep[e.g.,][]{2007ApJ...658..884C, Piorkowska:2013eww},
\begin{equation}
\frac{dn(\sigma_{\rm v},z_{l})}{d\ln \sigma_{v}}=n_{\rm z} \frac{\beta}{\Gamma(\alpha/\beta)}
\left(\frac{\sigma_{\rm v}}{\sigma_{\rm z}}\right)^{\alpha}\exp{\left[-\left(\frac{\sigma_{\rm v}}{\sigma_{\rm z}}\right)^{\beta}\right]},
\label{eq:lens}
\end{equation}
and
\begin{equation}
{n_{\rm z} = n_{0}(1+z)^{\kappa_{n}};\quad \sigma_{\rm z} = \sigma_{\rm v0}(1+z)^{\kappa_{v}}}
\end{equation}
where $\sigma_{\rm v0}$ is the characteristic velocity dispersion, $\alpha$ is the low-velocity power-law index, $\beta$ is the high-velocity exponential cutoff index, ${\Gamma(\alpha/\beta)}$ is the Gamma function, and 
$(n_{0}, \sigma_{\rm v0}, \alpha, \beta) = (0.008h^{3} {\rm Mpc}^{-3}, 161{\rm km \, s^{-1}}, 2.32, 2.67)$. The fitted evolution parameters $\kappa_{n}= -1.18$ and $\kappa_{v}=0.18$ are adopted from \citet{2021MNRAS.503.1319G}.  
%
%
\begin{figure}
    \centering
    \includegraphics[width=1.0\columnwidth]{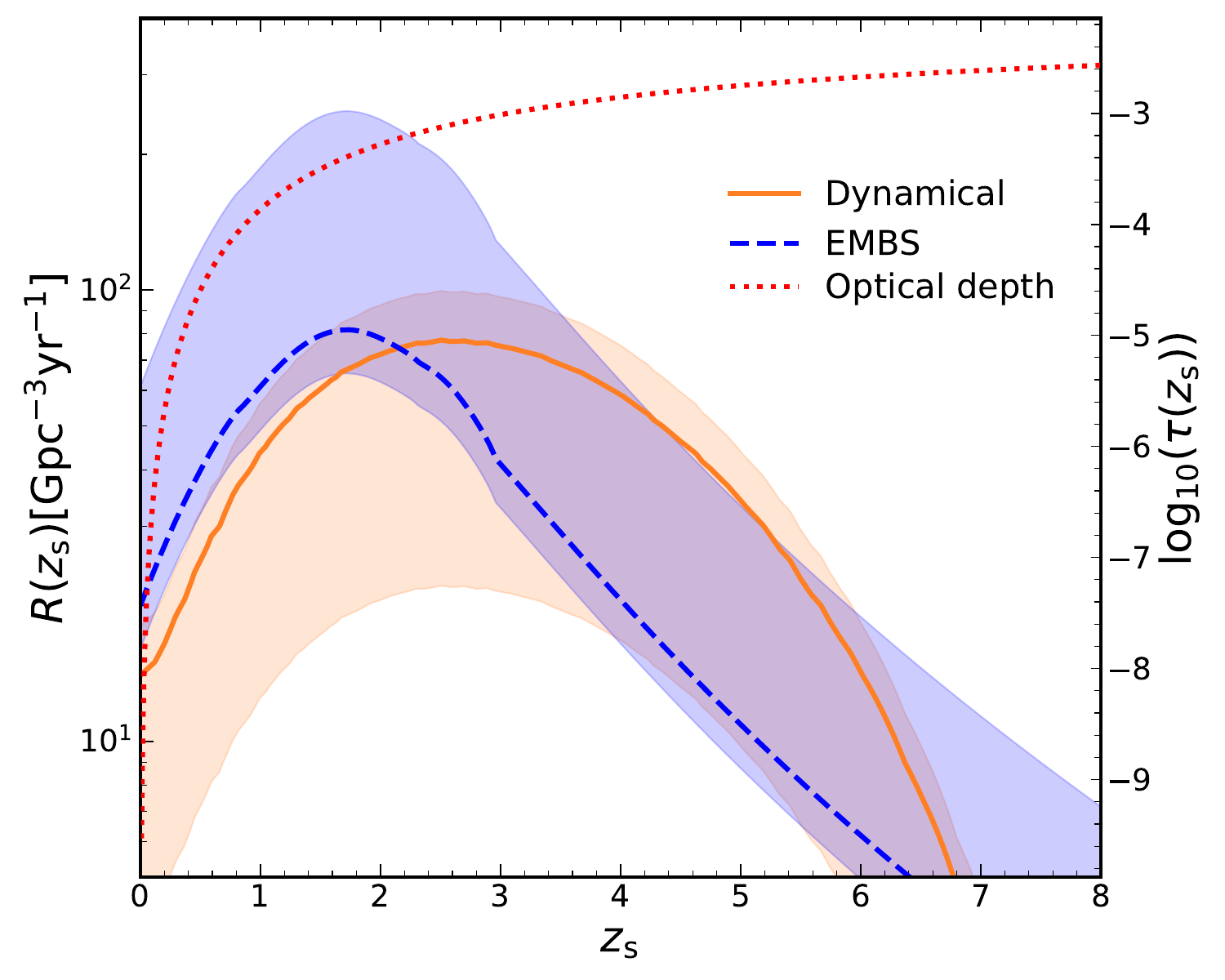}
    \caption{{Marginalize merger rate density and optical depth with redshift. The red dotted line represents the optical depth evolution $\tau({z_{\rm s}})$ for SIE + external shear model. The orange solid and blue dashed lines show the merger rate density evolution $R(z_{\rm s})$ scaled by the median estimated local merger rate density 
    of dynamical (see Sec\ref{subsec:rate}) and EMBS \citep[e.g.,][]{2017cao,2021MNRAS.500.1421Z,2022ApJ...940...17C} channel respectively. The corresponding shadows show the error induced by the uncertainties of local merger rate densities  {($90\%$ confidence interval)} \citep[e.g., ][]{2018ApJ...866L...5R,2021arXiv211103606T, 2021arXiv211103634T}.}}
    \label{fig:tau}
\end{figure}
The red-dotted line of figure \ref{fig:tau} shows the optical depth $\tau(z_{\rm s})$ evolution with redshift $z_{\rm s}$ calculated by equation~\ref{eq:tau} with the above approximations.  The higher the redshift of  the source is, the larger the probability to be lensed will be. 

\subsection{Monte Carlo method}
\label{sub:MC}

We apply the Monte Carlo simulation to obtain the mock lensed GW signals of sBBH mergers produced by dynamical channel with various parameters, including primary mass $m_1$, mass ratio $q$, primary spin $a_{z1}$, secondary spin $a_{z2}$ (we only consider the $z$-component for only the effective spin contributes to the GW signal waveform), redshift $z_{\rm s}$ and {the orientation angles} ($i$, $\theta$, $\phi$, $\psi$) \footnote{Here $\theta$ and $\phi$ are the declination (Dec) and right ascension (RA) of the GW source in the celestial coordinate system, while $i$ and $\psi$ give the source’s orientation with respect to the detector.}. 

The primary mass $m_1$, mass ratio $q$ and redshift $z_s$ of sBBH mergers are generated from the intrinsic number density $\dot{N}(z_s,m_1,q)$ discussed in Sec(\ref{subsec:rate}) with {Gibbs sampling method \citep[e.g., ][]{gibbs}}. We assume that $z_s$ is within $[0,10]$, $m_1$ is within $[5,85] M_{\odot}$ \citep[e.g., ][]{2021MNRAS.500.1421Z,2022MNRAS.511.5797M}, 
which is larger than that of sBBHs produced via EMBS channel.  We note here that GW events with redshift and primary mass beyond these intervals contribute very little to the total lensed rate for  $\dot{N}(z_s,m_1,q)$ declines rapidly in these regions.

Several works \citep[e.g.][]{2004PhRvD..70l4020S, 2016ApJ...832L...2R, 2017Natur.548..426F} have proposed that the spins of sBBHs formed by dynamical channel tend to be isotropic given the absence of a preferred direction and the persistence of an isotropic distribution through post-Newtonian evolution. Thus, we simply assume that the dimensionless spins $a_{1,2}$ are randomly and uniformly orientated and distributed within $[0,1]$, such that the distribution of  z-component spin, $a_{z}={a}\cdot {z}$, is a logarithmic form similar with the distribution in \citet{2017Natur.548..426F}. {Orientation angles, i.e., ($i$, $\theta$, $\phi$, $\psi$) are all uniformly and randomly sampled in the sky, which is almost consistent with the average numerical orientation probability density $\Theta$ proposed by \citet{1996PhRvD..53.2878F}. } 
{Noticed that in this work, the coalescence time is set to be $0$ and the coalescence phase is uniformly distributed.} We denote hereafter these parameters of the mock GW signal by vector ${\xi}=(a_1,a_2,i,\theta,\phi,\psi)$.

With the above parameters of GW signals, one may immediately calculate the detection rate of a lensed sBBH produced by the dynamical channel by equations~(\ref{source}), (\ref{eq:tau}), (\ref{eq:snr}), and (\ref{eq:mag}):
\begin{equation}
\begin{aligned}
    \dot{N}^{\rm len}(\rho^{\rm len} >\rho_{0}|z_{\rm s}) &=\int_{0}^{\infty} dz_{\rm s}\int_{0}^{\infty} dm_1\int_{0}^{1} dq \\
    & \times \int d\xi P(\rho^{\rm len} > \rho_{0} |\xi,z_{\rm s}, m_1,  q) \\
    & \times \dot{N}(z_{\rm s},m_1,q)\tau(z_{\rm s}) 
\end{aligned}
\end{equation}
%
%
where $P(\rho^{\rm len} > \rho_{0}|\xi,z_{\rm s}, m_1,  q)$ is the conditional probability of the $\rm SNR$ excess the threshold with a certain source parameter, i.e., $(\xi,z_{\rm s}, m_1,  q)$. The value of this probability should be either $1$ or $0$. 

\section{Results}
\label{sec:results}

\begin{table*}
\renewcommand\arraystretch{1.2}
    \caption{Prediction for the unlensed and lensed GW sBBH event rates  produced by dynamical or EMBS channel solemnly with various detectors. {$\dot{N}_s$ represent the total detectable GW sBBH event rates and $\dot{N}_{\ell}^{\rm d}$/ $\dot{N}_{\ell}^{\rm q}$ represent the lensed GW sBBH event rates with double/ quadruple images.  {Note here that the error of the detection rate is induced by the $90\%$ confidence uncertainty of the local merger rate density calibration and the detection threshold $\rho_{0}$ is set to be $8$. }} }
    \label{table:rate} 
    \resizebox{0.95\linewidth}{0.15\linewidth}{
    \begin{tabular}{lccccc}

    \hline 
    Detectors & Formation channels & $\dot{N}_{s}\left(>\rho_{0}\right)$ $\left(\mathrm{yr}^{-1}\right)$ & $\dot{N}_{\ell}^{\rm d}\left(>\rho_{0}\right)$ $\left(\mathrm{yr}^{-1}\right)$ & $\dot{N}_{\ell}^{\rm q}\left(>\rho_{0}\right)$ $\left(\mathrm{yr}^{-1}\right)$ \\
    \hline 
    
    ET & Dynamical & $2.6_{-1.9}^{+0.7} \times 10^{4}$ & $15.8_{-11.3}^{+4.52}$ & $0.52_{-0.37}^{+0.15}$  \\
    &EMBS & $2.5_{-0.4}^{+5.3} \times 10^{4}$ & $9.19_{-1.45}^{+20.3}$ & $0.30_{-0.05}^{+0.66}$  \\
    \hline
    CE & Dynamical & $3.2_{-2.3}^{+0.9} \times 10^{4}$ & $23.0_{-16.4}^{+6.56}$ & $0.79_{-0.56}^{+0.22}$  \\
    &EMBS & $2.8_{-0.4}^{+5.9} \times 10^{4}$ & $13.6_{-2.14}^{+30.0}$ & $0.45_{-0.07}^{+0.99}$ \\
    \hline
    ET+CE & Dynamical   & $3.2_{-2.3}^{+0.9} \times 10^{4}$ & $23.7_{-16.9}^{+6.76}$ & $0.85_{-0.61}^{+0.24}$                      \\
    & EMBS   &$2.8_{-0.4}^{+5.9} \times 10^{4}$ & $13.9_{-2.20}^{+30.7}$ & $0.47_{-0.07}^{+1.05}$                                \\
    \hline
    
    \end{tabular}}
    \begin{flushleft}
\end{flushleft}
\end{table*}

\begin{figure*}
    \centering 
    \includegraphics[width=0.8\columnwidth]{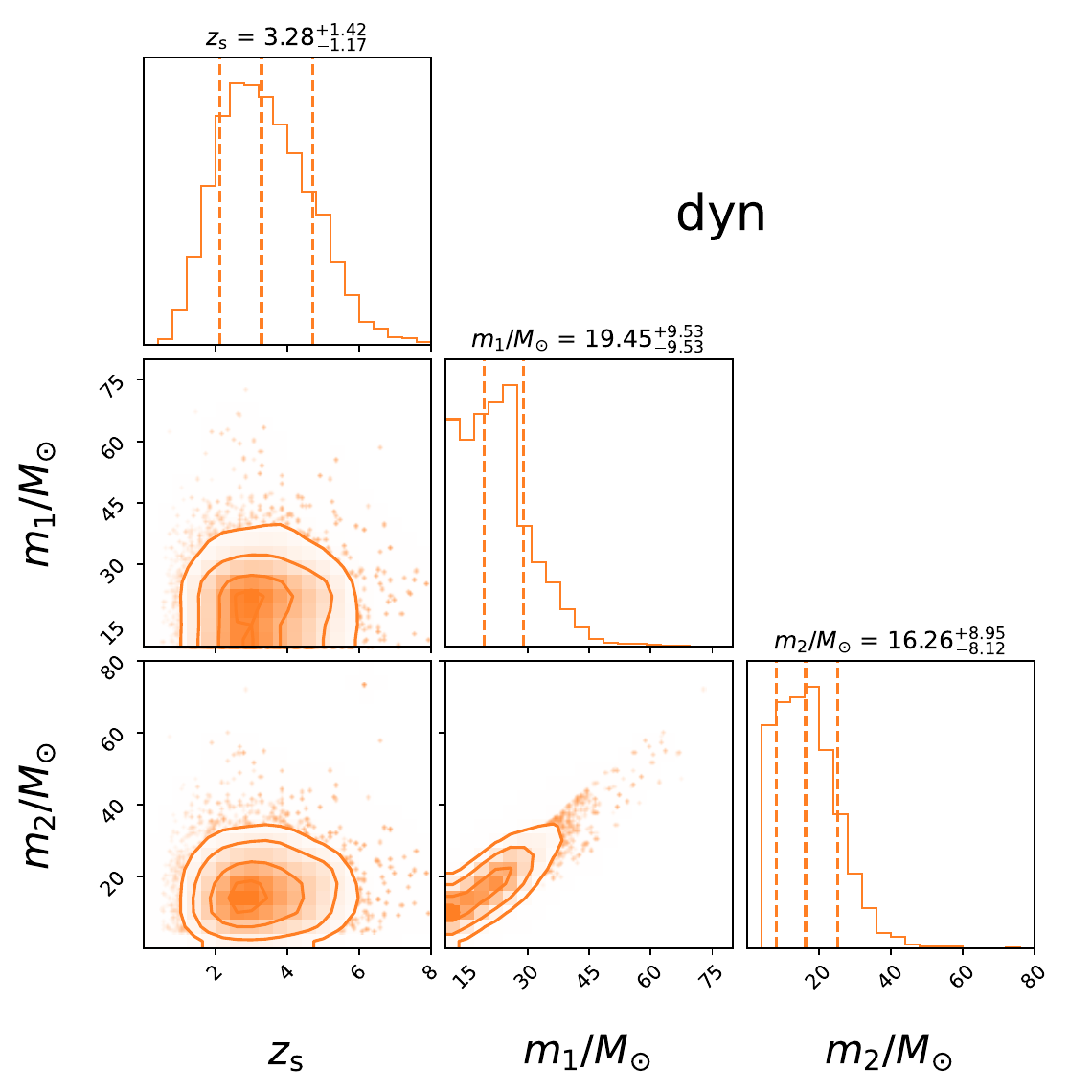}
    \includegraphics[width=0.8\columnwidth]{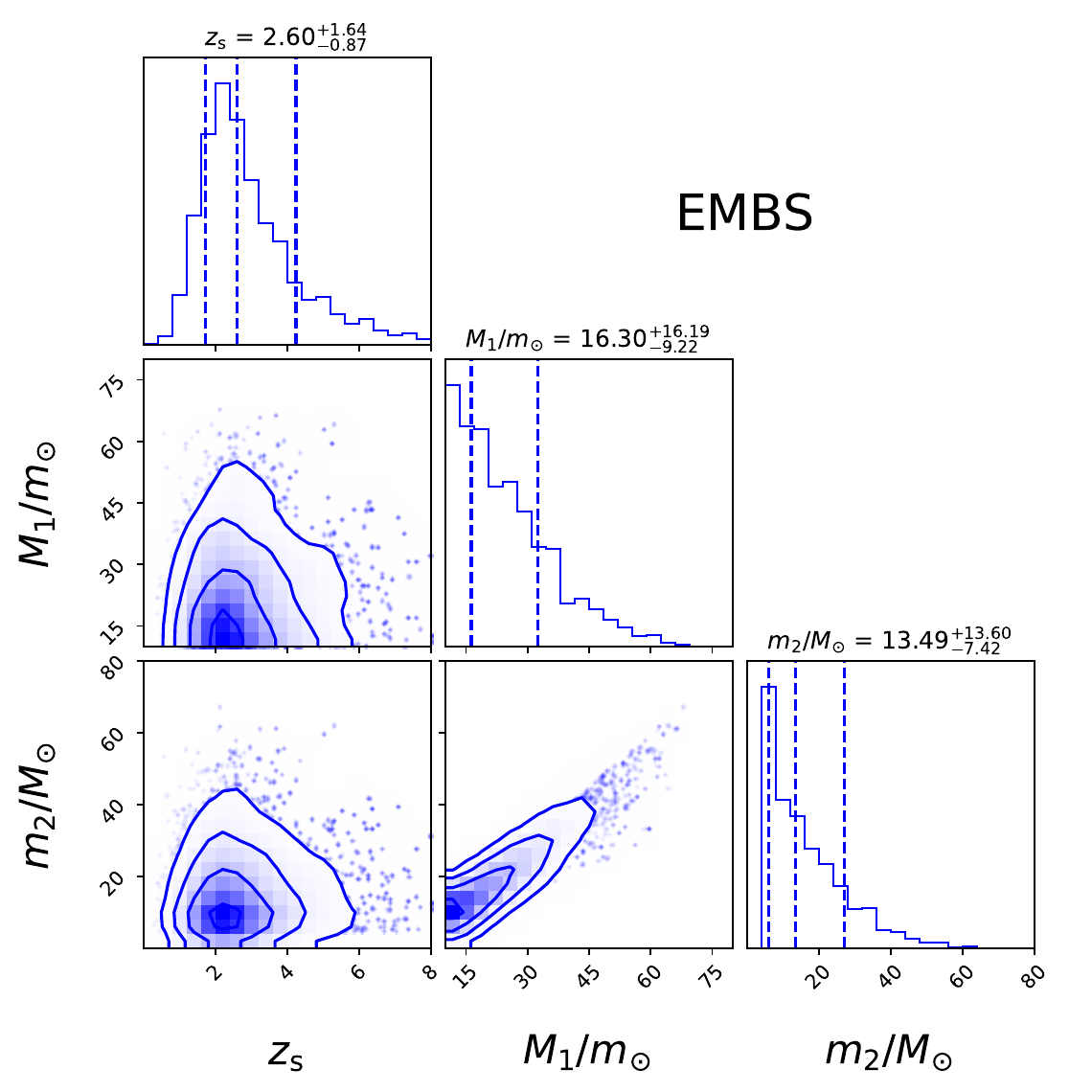}
    \caption{Redshift $z_{\rm s}$, primary mass $m_1$ and secondary mass $m_2$ distributions for lensed sBBH GW events, resulting from the dynamical and EMBS channels, respectively. The dashed lines and the solid contours indicate the $16\%$, $50\%$, and $84\%$ percentiles for each parameter.}
    \label{fig:dynamical}
\end{figure*}

\begin{figure}
    \centering
    \includegraphics[width=0.9\columnwidth]{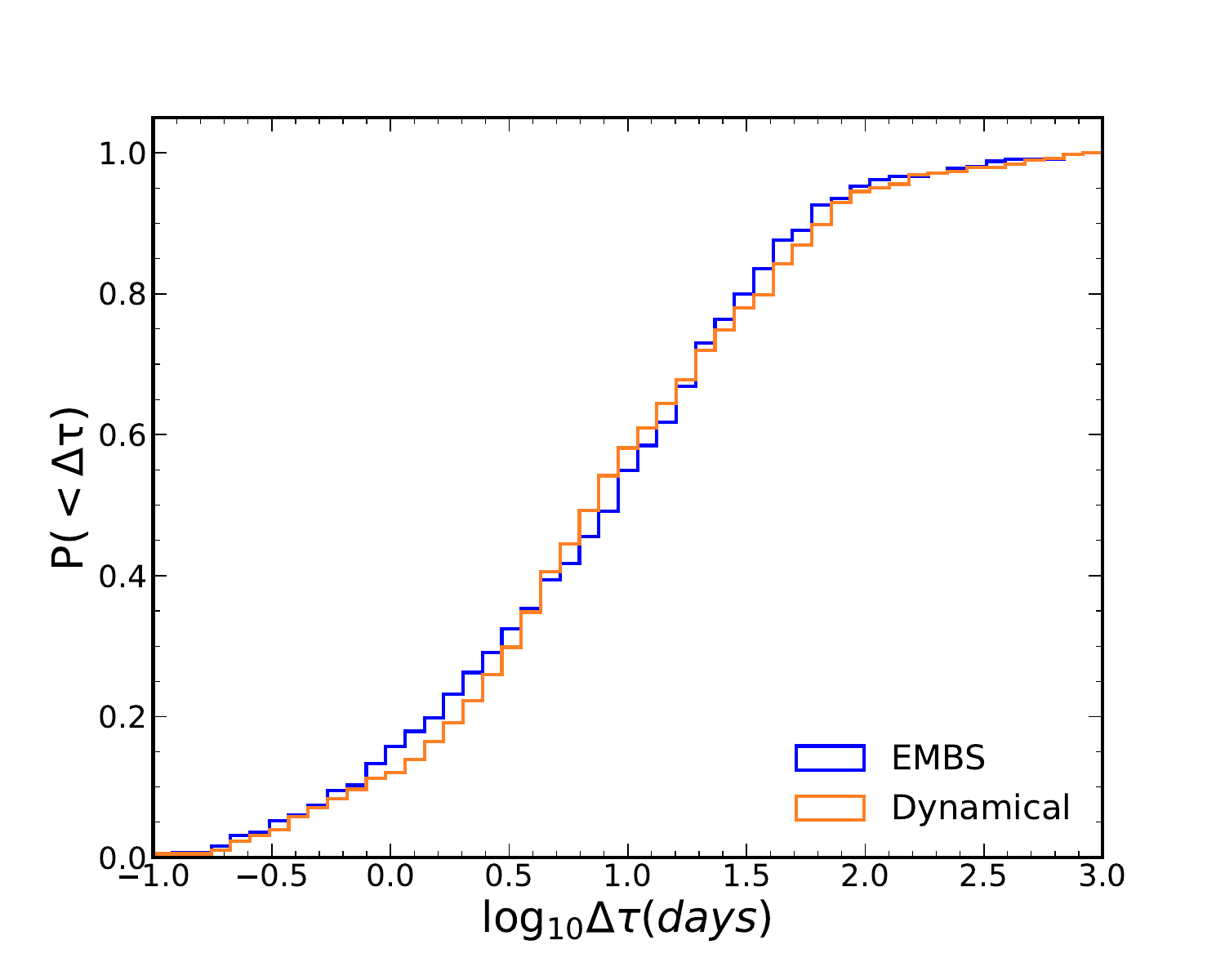}
    
    \caption{The cumulative time-delay distribution for the lensed sBBH mergers which have double images. The blue  and orange lines  show the results for EMBS and dynamical channel respectively.}
    \label{fig:time}
\end{figure}

Table \ref{table:rate} shows the predicted rate of unlensed and lensed detectable sBBHs, i.e., {$\dot{N}_s$ and $\dot{N}_l^{\rm d/q}$ } {(d denotes for the double images case and q denotes for the quadra images case)}, produced by dynamical channel with ET, CE and their networks. The uncertainty here is due to the uncertainties of the local merger rate densities, while the prediction rate adopts the median value respectively. For comparison, we also list the results of sBBHs originating from EMBS channel, using the same Monte Carlo procedures and a simple recipe for merger rate density analogous to \citet{2017cao} and \citet{2022ApJ...940...17C}. We note here that with the current design, CE is slightly more sensitive than ET and can detect almost all the sBBH events for their large masses. Therefore, there will be no significant promotion on the detection rates for no matter lensed or unlensed sBBHs with CE+ET networks compared to only individual CE detection, {if ignoring their different false alarm rate. }.

It can also be inferred that both the lensed rates for dynamical channels are significantly higher than those of the EMBS channel,  if adopting the median value of estimation on the local merger rate densities of both EMBS and dynamical channel respectively.  There are two reasons accounting for this discrepancy.
One is that the predicted intrinsic sBBH merger rate densities are different, due to different estimations on the local nerger rate density and their evolution with redshift. On the other hand, the intrinsic physical properties of sBBHs are strongly dependent on the formation channels.  {In Figure \ref{fig:tau}, we have shown that the lensing probability} (optical depth) increases more rapidly with redshift than the decreasement of $\rm SNR$ due to larger luminosity distance. Moreover, sBBH produced by the dynamical channel is more likely to have larger masses, which would enhance the GW $\rm SNR$. Thus, with the above two main contributors, the total lensed rate for sBBHs produced by the dynamical channel is $\sim 1.7$ times of that for the EMBS channel. {Here we have to notice that because of the large uncertainties on the local merger rate densities for both sBBH formation channels, the claims here may change choosing different local scale factors. However,  this ambiguilty may be resolved with the accumulation of detected sBBH merger GW events.}
\begin{figure}
    \centering
    \includegraphics[width=0.9\columnwidth]{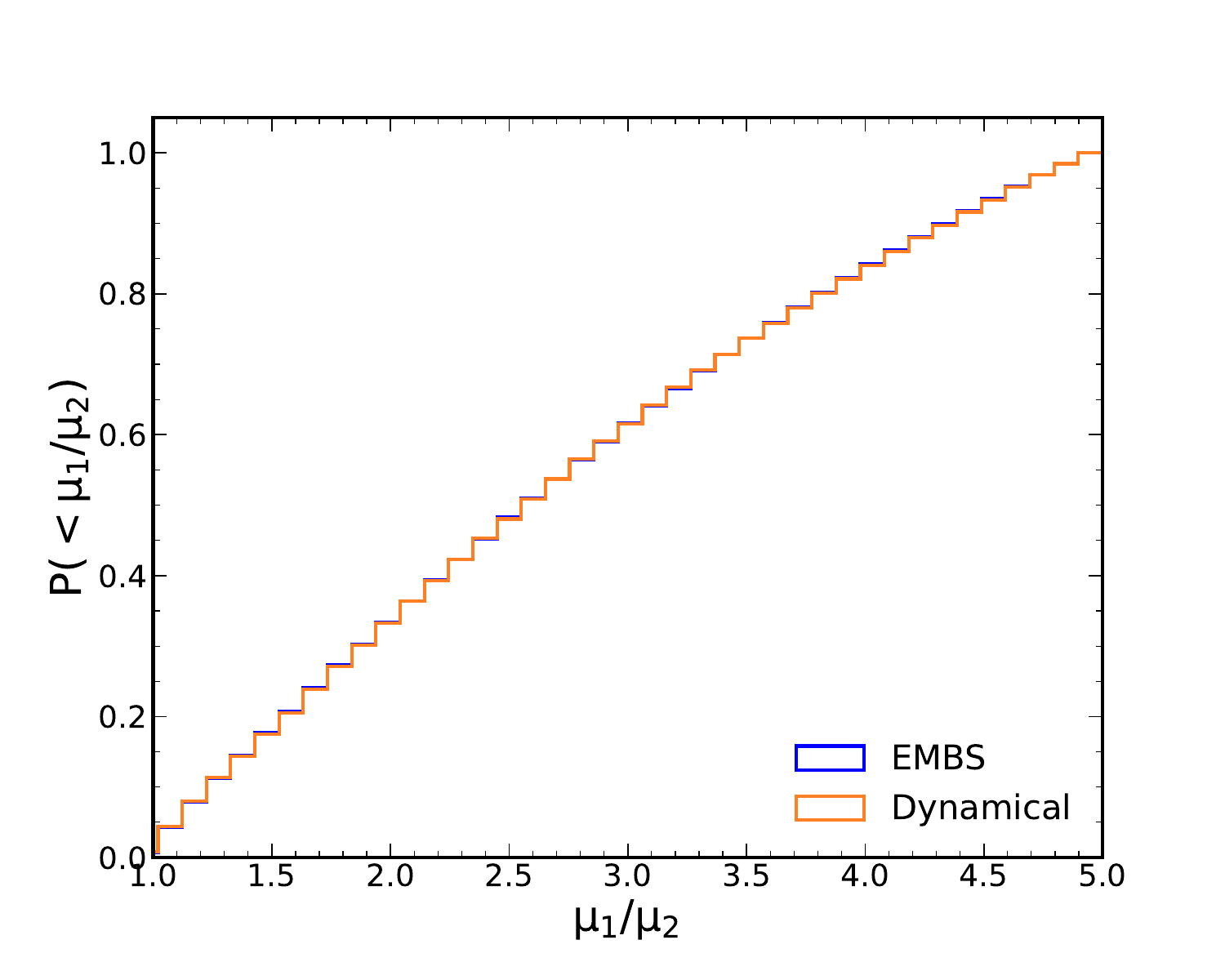}
    
    \caption{{The cumulative magnification ratio distribution for the lensed sBBH mergers which have double images. The blue and orange  lines show the results for EMBS and dynamical channel respectively.}}
    \label{fig:mag}
\end{figure}

Figure \ref{fig:dynamical} shows the redshift $z_{\rm s}$, primary mass $m_1$ and secondary mass $m_2$ distribution of sBBHs produced by dynamical (Left) and EMBS (Right) channels respectively. As seen from this figure, different sBBH formation channels may result in similar redshift distribution  (though the median is biased, $\sim 3.28$ for the dynamical channel and $\sim 2.60$ for EMBS channel), but significantly different distributions of primary mass and secondary mass. The peak of primary mass for the EMBS channel is about $7 \rm M_{\odot}$, while the peak for the dynamical channel is $\sim 30 \rm M_{\odot}$, which is mainly caused the detectability bias for GW detectors, i.e.,  sBBHs possessing small masses and lie far from detectors will have relatively low  $\rm SNR$, making them hard to be detected (see \ref{subsec:detect}). As shown in \ref{subsec:rate}, sBBHs formed by dynamical channels are much more distant but more massive than those formed by the EMBS channel. 
{We notice that there is large overlap in the parameter space of these two formation channels
for a specific lensed sBBH merger event. Nevertheless, one may  {constrain} their origins statistically for their distinct distribution of masses and redshifts.}

Figure \ref{fig:time} and \ref{fig:mag}  shows the cumulative distribution of time-delay and magnification ratio for the lensed sBBHs having double images. The dynamical channel tends to produce slightly more lensed sBBHs  with time-delay longer than $\sim 20$ days, which is partly caused by their higher redshifts compared to those from the EMBS channel.  {The magnification ratio is almost the same for both channels (varying between $1-5$), for the maginication factor is only dependent on the relative position of source and optical axis of the lens systems and independent of redshifts and other intrinsic parameters.}

{We note here that in the above results, we only consider single formation mechanisms for the GW sBBH sources, i.e., either dynamical or EMBS channels. This may not be the case in the real universe: both channels could contribute to the total detectable lensed rates. The total detection rate of the lensed GW sBBH is the sum of both channels, that is $\rm 25.0^{+24.8}_{-11.4}/ 36.6^{+36.5}_{-18.5}/ 37.6^{+37.5}_{-19.1}$ $\rm yr^{-1}$ for ET/CE/ET+CE repectively. If the local merger rate densities can be constrained more tightly, in principle, one may determine the composition of sBBHs in the
universe by checking the detectable lensed GW rates from sBBHs. We note here that this method  is strongly dependent on the merger rate density evolution models and their uncertainties for both the EMBS and dynamical channels. }

\section{Conclusions and discussions}

In this paper, we calculate the event rate of strongly lensed gravitational waves of stellar binary black hole mergers originating from the dynamical interactions in dense clusters. One may expect to detect $15.8/23.0/23.7$ (double images case, $0.52/0.79/0.85$ for quadruple images case) such events per year with third-generation ground-based GW detectors like ET, CE and ET+CE respectively, which is about twice larger than that produced by EMBS channel $\sim 9.19/13.6/13.9 \rm yr^{-1}$(double images case, $0.30/0.45/0.47$ for quadruple images case) for larger masses and luminosity distances. In addition, we also demonstrate that different compositions of sBBHs produced by the EMBS and dynamical channel will indeed affect the total predicted lensed rates, which may be a new approach to studying the origin of sBBHs.

We note here that there are many complexities one may take into account to make a more robust research. The merger rate density evolution of the dynamical channel varies with different numerical simulation results and therefore may change the predicted numbers of detectable lensed sBBH events. {As for the lensing statistics, we limit our paper to the galaxy-galaxy lensing, while ignoring the galaxy-cluster lensing simply because the cluster lensing is rarer \citep[e.g.,][the relative rate of lensed events by cluster is at most half of that estimated by galaxy-galaxy lensing]{2018MNRAS.475.3823S, 2022arXiv220412977S}.} We also simply assume the VDF of foreground galaxies follows a Schechter description, which only considers the elliptical galaxies. A mixed population of elliptical and spiral galaxies may affect the results slightly \citep[e.g.,][]{2014JCAP...10..080B}. 

{We also note that in this paper the detection rate is estimated without considering the false alarm probability (FAP) of GW detection on lensed events, which therefore can only be treated as an optimistic prediction. One way to estimate FAP is to 
calculate the Lens Bayes factor analytically \citep[e.g., ][]{2018arXiv180707062H,2023arXiv230413967G}. 
For example, \citet{2023arXiv230413967G} showed that $50.6\%$ of the lensed sBBH pairs detected by ET can be identified, while this number rises to $87.3\%$  for the CE+ET network, owing to the superior spatial resolution. Another way is to consider the parameter overlaps of those mock lensed pairs and impostors \citep{2023PhRvD.107f3023C}.  
We follow \citet{2023PhRvD.107f3023C} to make a qualitatively analysis here. The total FAP for the GW lensing detection is defined as the probability of at least one pair within a population of $\rm N$ events can mimic lensing due to astrophysical coincidence, i.e.,  
\begin{equation}
    \rm FAP= 1-(1-FAP_{pair})^{N_{pair}},
\end{equation}
where $\rm N_{pair}=N(N-1)/2$ is the total number of event pairs and $\rm FAP_{pair}$ is the percentage of these pairs with parameter overlaps that mimic the lensed pairs. As shown in Table \uppercase\expandafter{\romannumeral3} of \citet{2023PhRvD.107f3023C}, the combined $\rm FAP_{pair}$ is approximately the order of $10^{-5}\sim 10^{-6}$ for the $95\%$ confidence level lensed GW events detection of LVK network, which is the combination of false alarm due to mass, sky localization, and coalescence phase overlap. As for the future 3rd generation detectors, several works have predicted that these parameters may be constrained with precision by a factor of several ten times higher than (i.e., the overlap range in parameter space is smaller by several ten times ) than those by the current LVK observations, beneficial from the substantially higher sensitivity of CE and ET \citep[e.g., ][]{2018PhRvD..97f4031Z,2022NatSR..1217940P}, which may reduce $\rm FAP_{pair}$ by a order of $10^{-3}-10^{-4}$. Then the total $\rm FAP$ for the lensed GW events detected by the 3rd generation detectors is on the order of $10^{-2}-10^{0}$ per year, which is significantly smaller than the predicted number of the detection rate shown in Table\ref{table:rate}. Therefore, we are optimistic on the detection of lensed GW events in the coming new era of 3rd generation ground-based detectors.}

 {Once the corresponding lensed host galaxies could be observed at the same time}, it is possible to locate the relative positions of sBBHs in host galaxies by mapping time-delay and magnification factors \citep[e.g.][]{2020MNRAS.498.3395H, 2022arXiv220408732W,2022ApJ...940...17C}, which provide a new probe to understand the origin of sBBHs according to their spatial distributions in host galaxies. Thus, it is crucial to study the detectability of {these both lensed events}. With the same settings analogous to \citet{2022ApJ...940...17C}, one can multiply the detection rate of lensed sBBHs proposed in this paper to estimate
the detection rate of both lensed events produced by dynamical channels with future sky-surveys. That is $0.8^{+0.1}_{-0.5} \rm yr^{-1}$ for CSST and  $0.6^{+0.2}_{-0.4} \rm yr^{-1}$ for Euclid with  ET+CE networks if assuming all lensed sBBHs are produced by dynamical channel . Noticed that if considering the reconstruction
errors for the lensed host galaxies and the sky localization errors of the GW signals, these rates may reduce by a factor of $\sim 0.35-0.20$ according to \citet{2022arXiv220408732W} with the assumption of $1-5$ deg$^2$ sky
localization errors.

\label{sec:con}

\section*{acknowledgement}
We thank the referee for his/her careful reading and insightful comments. We also thank Professor Youjun Lu, Hao Ma and Yuetong Zhao for their insightful discussions and helpful suggestions.
This work is partly supported by the National Natural Science Foundation of China (Grant No. 12273050, 11690024, 11873056, 11991052), the Strategic Priority Program of the Chinese Academy of Sciences (Grant No. XDB 23040100), and the National Key Program for Science and Technology Research and Development (Grant No. 2020YFC2201400 and 2016YFA0400704).

\bibliographystyle{aasjournal}
\bibliography{ref.bib}

\end{document}